\documentclass[conference]{IEEEtran}
\usepackage{comment}
\usepackage{graphicx}
\usepackage{amssymb}
\usepackage{mathtools}
\usepackage{dsfont}
\usepackage{cite}
\usepackage{stfloats}
\usepackage{subfigure}
\usepackage{psfrag}
\usepackage[mathscr]{euscript}
\usepackage{acronym}  % make an acronym
\usepackage{amsmath}
\usepackage{algorithm}
\usepackage[noend]{algpseudocode}
\usepackage{booktabs}
\usepackage{bbm}

\makeatletter
\def\BState{\State\hskip-\ALG@thistlm}
\makeatother

\usepackage{float}
\usepackage{balance}

\acrodef{CCDF}{complementary cumulative distribution function}
\acrodef{CF}{characteristic function}
\acrodef{PPP}{Poisson point processe}
\acrodef{RV}{random variable}
\acrodef{i.i.d.}{independent and identically distributed}
\acrodef{PDF}{probability distribution function}
\acrodef{CDF}{cumulative distribution function}
\acrodef{ch.f.}{characteristic function}
\acrodef{AWGN}{additive white Gaussian noise}
\acrodef{SNR}{signal-to-noise ratio}
\acrodef{LRT}{likelihood ratio test}
\acrodef{DRT}{distance ratio test}
\acrodef{GLRT}{generalized likelihood ratio test}
\acrodef{CRLB}{Cram\'{e}r-Rao lower bound}
\acrodef{CRB}{Cram\'{e}r-Rao bound}
\acrodef{ZZLB}{Ziv-Zakai lower bound}
\acrodef{ZZB}{Ziv-Zakai bound}
\acrodef{LOS}{line-of-sight}
\acrodef{ToF}{time-of-flight}
\acrodef{NLOS}{non-line-of-sight}
\acrodef{GDOP}{geometric dilution of precision}
\acrodef{GPS}{Global Positioning System}
\acrodef{FIM}{Fisher information matrix}
\acrodef{PEB}{position error bound}
\acrodef{SPEB}{squared position error bound}
\acrodef{TOA}{time-of-arrival}
\acrodef{TOF}{time-of-flight}
\acrodef{WSN}{wireless sensor network}
\acrodef{MAC}{medium access control}
\acrodef{RSS}{received signal strength}
\acrodef{WAF}{wall attenuation factor}
\acrodef{TDOA}{time difference-of-arrival}
\acrodef{RF}{radiofrequency}
\acrodef{RTT}{round-trip time}
\acrodef{AOA}{angle-of-arrival}
\acrodef{MF}{matched filter}
\acrodef{ED}{energy detector}
\acrodef{ML}{maximum likelihood}
\acrodef{MSE}{mean-square error}
\acrodef{RMSE}{root-mean-square error}
\acrodef{LEO}{localization error outage}
\acrodef{ppm}{part-per-million}
\acrodef{ACK}{acknowledge}
\acrodef{UWB}{Ultrawide bandwidth}
\acrodef{TNR}{threshold-to-noise ratio}
\acrodef{LS}{least squares}
\acrodef{IR-UWB}{impulse radio UWB}
\acrodef{FCC}{Federal Communications Commission}
\acrodef{TH}{time-hopping}
\acrodef{PPM}{pulse position modulation}
\acrodef{MUI}{multi-user interference}
\acrodef{PDP}{power delay profile}
\acrodef{BPZF}{band-pass zonal filter}
\acrodef{SIR}{signal-to-interference ratio}
\acrodef{SINR}{signal-to-interference-plus-noise ratio}
\acrodef{RFID}{radio frequency identification}
\acrodef{WPAN}{wireless personal area network}
\acrodef{WWB}{Weiss-Weinstein bound}
\acrodef{DP}{direct path}
\acrodef{MF}{matched filter}
\acrodef{MMSE}{minimum-mean-square-error}
\acrodef{SBS}{serial backward search}
\acrodef{SBSMC}{serial backward search for multiple clusters}
\acrodef{NBI}{narrowband interference}
\acrodef{WBI}{wideband interference}
\acrodef{INR}{interference-to-noise ratio}
\acrodef{CR}{channel response}
\acrodef{CIR}{channel impulse response}
\acrodef{CR}{channel  response}
\acrodef{RADAR}{radar}
\acrodef{MUR}{Multistatic radar}
\acrodef{JBSF}{jump back and search forward}
\acrodef{HDSA}{high-definition situation-aware}
\acrodef{RRC}{root raised cosine}
\acrodef{ST}{simple thresholding}
\acrodef{BTB}{Bellini-Tartara bound}
\acrodef{P-Max}{$P$-Max}  %suggestion, use with \acl{P-Max}
\acrodef{MIMO}{multiple-input multiple-output}
\acrodef{MAP}{maximum a posteriori}
\acrodef{FG}{factor graph}
\acrodef{OP}{outage probability}
\acrodef{WED}{wall extra delay}
\acrodef{RMS}{root mean square}
\acrodef{SPAWN}{sum-product algorithm over a wireless network}
\acrodef{MDD}{minimum distance distribution}
\acrodef{MAP}{maximum a posteriori probability}
\acrodef{SAP}{small cell access point}
\acrodef{UE}{user equipment}
\acrodef{MBS}{macro cell base station}
\acrodef{UER}{\ac{UE} Relay}
\acrodef{D2D}{device-to-device}
\acrodef{MBS}{macro base station}
\acrodef{CSI}{channel state information}
\acrodef{OGR}{outage guard region}
\acrodef{FUR}{feasible UER region}
\acrodef{EHR}{energy harvesting region}
\acrodef{EH}{energy harvesting}
\acrodef{D2D-EHSN}{D2D communication provided \ac{EH} small cell network}
\acrodef{D2D-EHHN}{D2D communication provided \ac{EH} heterogeneous network}
\acrodef{3GPP}{3rd Generation Partnership Project}
\acrodef{BS}{base station}
\acrodef{DF}{decode and forward}
\acrodef{CCDF}{complementary cumulative distribution function}
\acrodef{ZF}{zero forcing}
\acrodef{RZF}{regularized zero forcing}
\acrodef{WLLN}{weak law of large number}
\acrodef{SLLN}{strong law of large numbers}
\acrodef{TDD}{Time-division duplex}
\acrodef{EE}{energy efficiency} 
\acrodef{HetNet}{heterogeneous network} 
\acrodef{SCP}{Single Cell Processing}
\acrodef{CBF}{Coordinated Beamforming}
\usepackage{color}
\usepackage{dsfont}
\usepackage{bbm}

\DeclareMathAlphabet{\mathsf}{OML}{cmbr}{m}{it}
\newcommand{\red}[1]{{\textcolor[rgb]{1,0,0}{#1}}}

\newtheorem{theorem}{\bf Theorem}
\newtheorem{lemma}{\bf Lemma}

\newcommand{\bd}{\begin{description}}
\newcommand{\ed}{\end{description}}
\newcommand{\be}{\begin{enumerate}}
\newcommand{\ee}{\end{enumerate}}
\newcommand{\bi}{\begin{itemize}}
\newcommand{\ei}{\end{itemize}}
\newcommand{\bl}{\begin{list}}
\newcommand{\el}{\end{list}}
\newcommand{\bt}{\begin{tabbing}}
\newcommand{\et}{\end{tabbing}}

\newlength \figwidth
\newcommand{\paperTitle}{Analysis of Age of Information in\\Non-terrestrial Networks}

\begin{document}

% This code is to reduce the list of authors by using et. al:
\bstctlcite{IEEEexample:BSTcontrol}

\title{\paperTitle}

\author{%
  \IEEEauthorblockN{Yanwu~Lu\IEEEauthorrefmark{2},
                    Howard~H.~Yang\IEEEauthorrefmark{2},
                    Nikolaos Pappas\IEEEauthorrefmark{3},
                    Giovanni Geraci\IEEEauthorrefmark{1},
                    Chuan Ma\IEEEauthorrefmark{4},
                    and Tony Q. S. Quek\IEEEauthorrefmark{5}}\vspace{0.2cm}
  \IEEEauthorblockA{\IEEEauthorrefmark{2}%
                    ZJU-UIUC Institute, Zhejiang University, Haining, China}
                    %Zhejiang University, 718 Haining, China,
  \IEEEauthorblockA{\IEEEauthorrefmark{3}%
                    Department of Computer and Information Science, Link\"{o}ping University, Link\"{o}ping, Sweden}
\IEEEauthorblockA{\IEEEauthorrefmark{1}%
                    Telefonica Research and Universitat Pompeu Fabra, Barcelona, Spain}
\IEEEauthorblockA{\IEEEauthorrefmark{4}%
                    Zhejiang Lab, Hangzhou, China
                    } 
\IEEEauthorblockA{\IEEEauthorrefmark{5}%
                    Information System and Technology Design Pillar, Singapore University of Technology and Design, Singapore} 
                    
                    %Link\"{o}ping University, 581 83 Link\"{o}ping, Sweden,
                    % chi-ling@zju.edu.cn,\{haoyang, mengzhang\}@intl.zju.edu.cn, nikolaos.pappas@liu.se}
}

\maketitle

\thispagestyle{empty}

\begin{abstract}

Non-terrestrial networks (NTN), particularly low Earth orbit (LEO) satellite networks, have emerged as a promising solution to overcome the limitations of traditional terrestrial networks in the context of next-generation (6G) wireless systems. In this paper, we focus on analyzing the timeliness of information delivery in NTN through the concept of Age of Information (AoI). We propose an on-off process to approximate the service process between LEO satellites and a source node located on the Earth's surface. By utilizing stochastic geometry, we derive a closed-form expression for the time-average AoI in an NTN. This expression also applies to on-off processes with one component following an exponential distribution while the other has its probability density function supported on a bounded interval. Numerical results validate the accuracy of our analysis and demonstrate the impact of source status update rate and satellite constellation density on the time-average AoI. Our work fills a gap in the literature by providing a comprehensive analysis of AoI in NTN and offers new insights into the performance of LEO satellite networks.

\begin{comment}
    \red{We investigate the age of information (AoI) in the context of non-terrestrial communications. 
    We consider a pair of source-destination nodes located on the earth and separated by long distance. 
    The source node sends information packets, containing its latest status updates, to the destination, where the communication take place over the satellites. 
    Due to satellites' roaming nature, the availability of non-terrestrial connections to the source node is temporal and constitutes an on-off process in time, the transmissions during the off periods are considered to be lost.
    We leverage tools from stochastic geometry to characterize the statistical property of such a process and derive a closed-form expression for the time-average AoI.
    The accuracy of our analysis is verified by simulations. }
\end{comment}

\end{abstract}

%
% \begin{IEEEkeywords}

% \end{IEEEkeywords}

% ============================================ %
%         Section: Introduction                %
% ============================================ %
\section{Introduction}\label{sec:intro}

% \cite{GerLopBen2023,giordani2020non,LinRomRul2021,DarKurYan2022,BenGerLop2022,RinMaaTor2020,SedFelLin2020,kodheli2020satellite,GuiVanCon2019,AlhAlhWan2022,LeySorMat2022,LinCioCha2021}

Non-terrestrial networks (NTN) have emerged as a potential technological breakthrough in the realm of wireless systems, particularly in the context of next-generation (6G) networks \cite{LinRomRul2021,giordani2020non,LinCioCha2021,GerGarAza2022,OugGerPol2023}. NTN offer a promising solution to overcome the limitations of traditional terrestrial networks \cite{GerLopBen2023,WanGerQue2023,BenGerLop2022}. Mega satellite constellation design and standardization efforts are currently underway \cite{LeySorMat2022,DarKurYan2022}, showcasing the growing interest in this field. Among NTN platforms, low Earth orbit (LEO) satellites have gained significant attention due to their advantages of relatively low latency and operating cost \cite{AlhAlhWan2022,kodheli2020satellite}. Numerous LEO satellites have already been deployed at various orbits to provide global commercial communication services \cite{del2019technical}. However, as the number of deployed LEO satellites increases, the performance of the network, particularly the timeliness of information delivery, requires careful investigation.

In order to assess the timeliness of information successfully received in an NTN, the concept of \emph{Age of Information} (AoI) may serve as an appropriate metric \cite{sunmodiano2019age, pappas2023age}. AoI provides insights into the freshness of information by measuring the time elapsed since the generation of the most recent update at a particular destination \cite{kadota2018optimizing}. By analyzing the mean AoI, it may become possible to estimate the delivery performance of the entire network \cite{9380899}. It is worth noting that LEO satellites operate at extremely fast speeds in orbit, resulting in frequent transitions between connected and disconnected states. These transitions inevitably lead to an increase in AoI.

To analyze the performance of NTN, researchers have employed stochastic geometry as a mathematical tool, enabling the modeling of satellite locations and the derivation of expressions for coverage probability \cite{jung2022performance}. For instance, previous studies have modeled the location of LEO satellites using a binomial point process (BPP) on a sphere \cite{talgat2020stochastic}. Other works have considered systems of multiple concentric spheres, with satellites uniformly distributed as BPP on each sphere \cite{talgat2020nearest}. Analytic formulations for coverage probability in dense satellite constellations have also been proposed \cite{al2021analytic,okati2020downlink}. 
% Additionally, the motion and service processes between satellites and the ground have been studied, considering scenarios such as source node movement at a fixed velocity \cite{madadi2017shared} and on-off service processes \cite{sinha2022age}.

While these existing works have individually addressed certain aspects of NTN, there remains a gap in the literature when it comes to analyzing the AoI in such networks. 
Specifically, owing to the various positions of satellites and their communication capabilities, the connected (and disconnected) periods of the satellite links are strongly affected by network configurations. The AoI performance has not yet been well understood in this scenario.
% the stochastic behavior of the transitions between connected and disconnected states associated with the satellite-source node connection has yet been characterized. 
% Additionally, the AoI performance in an NTN has yet been well explored. 
% (As high latency is one shortcoming of NTN, AoI is better than other aspects to provide a mathematical tool to optimize this problem, and analyzing the AoI under the on-off process approximated by the operation of NTN might be challenging.)
In this article, we aim to fill this gap by studying the AoI in a LEO satellite network.
%
% \footnote{\gio{Is this a straightforward extension? If not, why?}}
%
We consider a system where satellite positions follow a Poisson point process (PPP) and model the satellite service process as an on-off process, where updates arriving during the off-service period are dropped. 
The recent study \cite{sinha2022age} has analyzed AoI under an on-off process, and is closely related to our work. However, the analysis in \cite{sinha2022age} was limited to the case of on and off periods being exponentially distributed, which, is unable to express satellite service process in this paper.
Our main contributions can be summarized as follows:
\begin{itemize}
\item We formulate, for the first time, the notion of AoI in an NTN. We do so by establishing an on-off process that approximates the service process between orbiting LEO satellites and a source node located on the Earth.
% end-users.
%
\item
Through stochastic geometry tools, we derive a closed-form expression for the time-average AoI in an NTN. { This result also extends the AoI analysis under on-off processes \cite{sinha2022age} into scenarios where one component follows an exponential distribution while the other has its probability density function supported on a bounded interval.}
\item We provide numerical results that validate the accuracy of our analysis and quantify the impact of both the source status update rate and the satellite constellation density on the time-average AoI.
\end{itemize}

\section{System Model}\label{sec:sysmod}
% The whole system model includes two parts, one is about the composition and the operation mode of the network, and the other is to define the performance metric.

\begin{figure}[t!] 
  \centering{}

    {\includegraphics[width=0.75\linewidth]{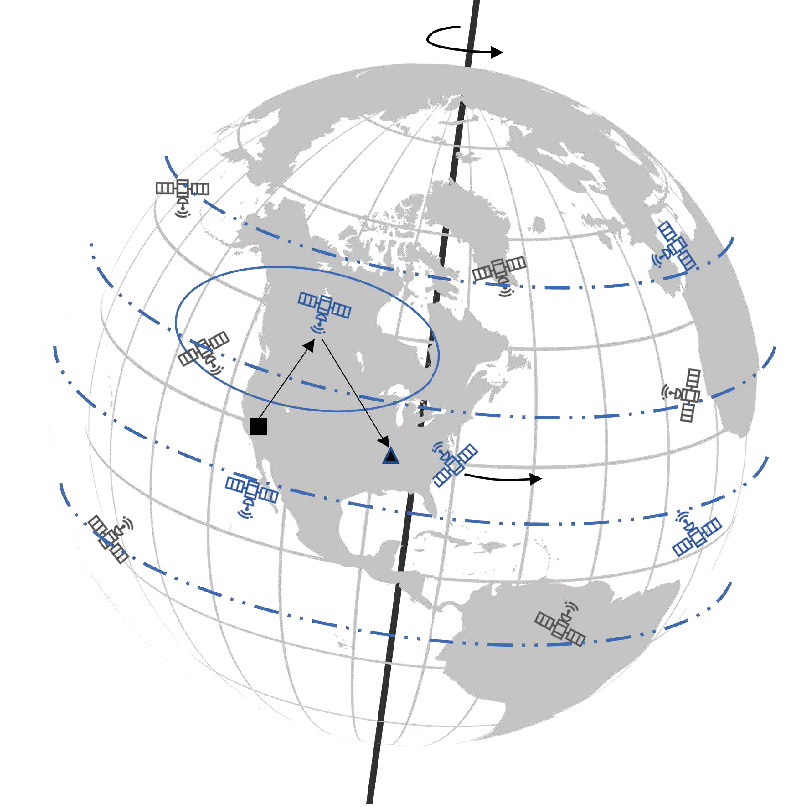}}
    \vspace{-4mm}
  \caption{Illustration of the NTN under consideration, with a constellation of LEO satellites deployed on a sphere and connecting source-destination nodes.}
  \label{fig:Sat}
\end{figure}

In this section, we introduce the network topology under consideration, the traffic generation and propagation channel models, and the performance metric of our analysis.

\subsubsection*{Network topology}
We consider an NTN consisting of a constellation of LEO satellites deployed at the same altitude $h$. We assume the positions of the satellites to follow a homogeneous Poisson point process (PPP) \cite{al2021analytic, al2021optimal, park2022tractable} of intensity $\lambda$ on a sphere of radius $R_\oplus+h$, where $R_\oplus = 6371\,\mathrm{km}$ denotes the Earth radius. 
We focus on a source node on the ground that updates its latest status to a destination node. We consider the scenario where the source and destination are both out of the coverage of a terrestrial network. Hence, communication needs to take place via an NTN, through satellite nodes. An illustrative example of this setup is provided in Fig.~\ref{fig:Sat}.

\subsubsection*{Traffic generation and propagation channel}
We assume the source node to employ a \textit{generate-at-will} approach \cite{pappas2023age} for status updates, where a packet is transmitted immediately after its generation.  
Specifically, we consider the interval between the generation of two consecutive status updates to be independently and identically distributed (i.i.d.), following an exponential distribution with rate $\mu$. 
We further assume the source node to send out information packets at a fixed transmit power $P_{\mathrm{tx}}$, where the signal propagation is subject to path loss that obeys power law with path loss exponent $\alpha$, and reception experiences white Gaussian noise with variance $\sigma^2$. 

\subsubsection*{Coverage and information delivery}
We deem a source node to be covered by a satellite when its signal-to-noise ratio (SNR) surpasses a minimum decoding threshold $\theta$. Out of the satellites from which it receives coverage, the source node then connects to at most one, namely the one providing the largest received power. 
If the source node is connected to a satellite upon generating a new status update, the corresponding information packet can be successfully delivered to the destination after experiencing a constant propagation delay $D$.%
\footnote{We assume the transmission from a satellite to the destination node to be always available. We focus on the uplink transmission from the source node to a satellite as this typically represents the bottleneck due to a limited transmission power budget.
}
However, owing to the mobility nature of satellites, any satellite providing coverage eventually moves away from the source node, being replaced by a new satellite. 
Before the arrival of another satellite, the source node will remain out of coverage, with any status updates generated during this period being undelivered. 

% Additionally, we do not consider re-transmission and/or acknowledgment of reception because of the non-negligible propagation latency. 
% As such, every information packet will only be sent out once only.

%%%%%%%%%%%%%%%%%%%%%%%%%%%%%%%%%%%%%%%%%%%%%%%%%%%

\begin{figure}[t!] 
  \centering{}
    {\includegraphics[width=0.9\figwidth]{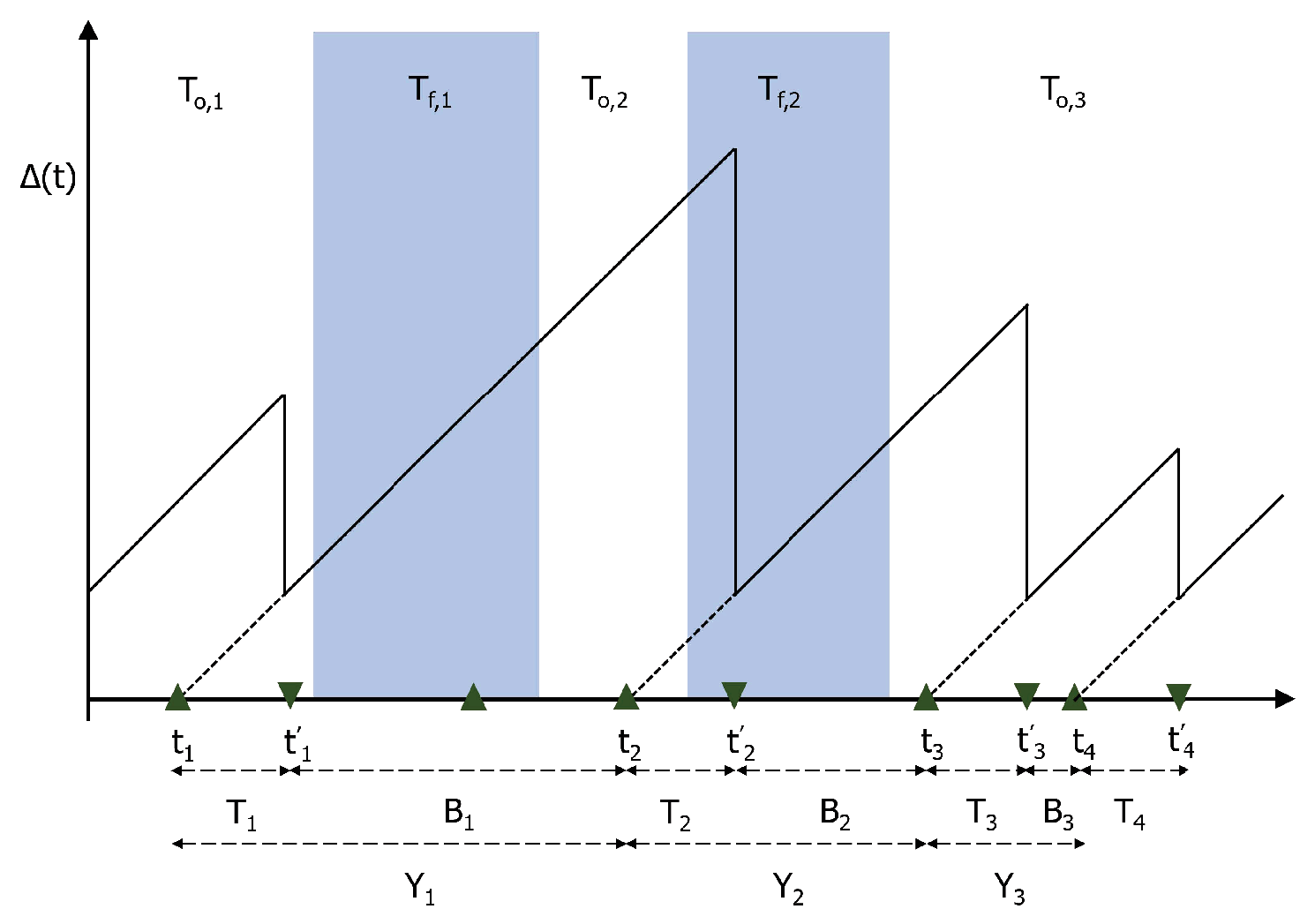}}
  \caption{Example of AoI evolution under on- and off-service periods, which is relevant to intermittent NTN connectivity.}
  \label{fig:AoI}
\end{figure}
\subsubsection*{Key performance indicator}

In this paper, we study the AoI in an NTN. AoI quantifies the \emph{freshness} of the information received at a destination node, and it is formally defined as follows:
\begin{align} \label{equ:AoI}
    \Delta (t) = t - G(t),
\end{align}
where $G(t)$ is the timestamp at which the latest update received by the destination at time slot $t$ was generated at the source. 
The inherent dynamics of the NTN cause the availability of satellite connectivity to be intermittent. Therefore, updates from the source node can only be successfully delivered while such connectivity is available. As a result, the evolution of the AoI in an NTN follows a trajectory as per Fig.~\ref{fig:AoI}, where shaded areas correspond to intervals of lack of connectivity, during which any newly generated updates are lost. To capture the overall timeliness in the delivery of status updates through an NTN, we then define the time-average AoI as follows:
\begin{align}
    \bar{\Delta} = \lim_{T \to \infty }{\frac{1}{T}\int_{0}^{T} \Delta(t)}.
\end{align}

In the next section, we derive an expression for $\bar{\Delta}$ in the NTN under consideration, and we shed light on how it is affected by the key NTN system-level parameters.

\section{Analysis of AoI in NTN}\label{sec:age_analysis}

We begin by characterizing the distribution of the intervals during which the source node is connected/disconnected to/from the NTN, which we denote as the \emph{on}- and \emph{off-service} periods. We then derive the analytical expression for the time-average AoI and discuss special cases to provide additional insights. 

\subsection{Distribution of the On-Off Service Periods}

\begin{figure}[t!] 
  \centering{}
  \subfigbottomskip = 25pt
  \subfigure [Side view of the NTN, with a shaded area containing satellites whose distance from the source node is smaller than $r_\mathrm{max}$.]{
  \includegraphics[width=\figwidth]{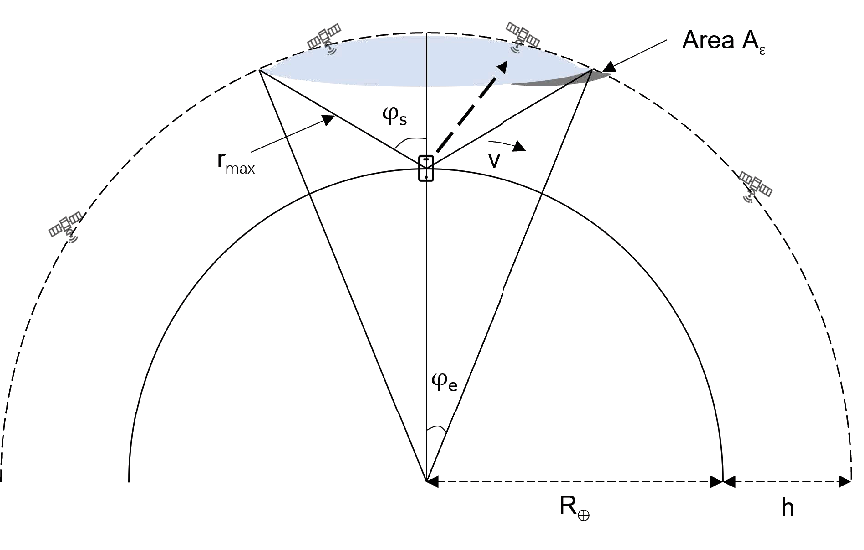}
  \label{fig:sketch}}
  \subfigure[NTN as seen from the standpoint of the source node, with a shaded area indicating the relative movement of the satellites with respect to the source node.]{
  \includegraphics[width=\figwidth]{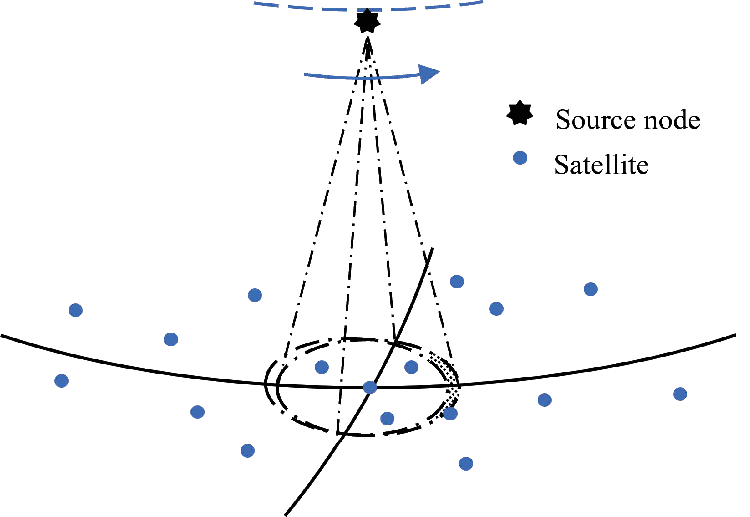}
  \label{fig:user}}
  \vspace{-8mm}
  \caption{Illustration of the relative motion between a source node and satellites.}
\end{figure}

At time $t$, let $r(t)$ denote the distance between the source node and the closest satellite. Then the corresponding SNR can be expressed as
\begin{align} \label{equ:SNR_exprss}
    \text{SNR}(t) = \frac{P_{\mathrm{tx}} \, r(t)^{-\alpha}}{\sigma^2}.
\end{align}

Since the source node can reliably connect to the satellite only when the SNR surpasses a minimum decoding threshold, i.e., $\text{SNR}(t)>\theta$, equation \eqref{equ:SNR_exprss} yields a maximum transmission distance for successful decoding, $r_\mathrm{max} = (P_{\mathrm{tx}}/{\sigma^2\theta})^{\frac{1}{\alpha}}$. 
From the perspective of the source node, the maximum transmission distance separates the sphere over which the satellites are located into two parts, as depicted in Fig.~\ref{fig:sketch}: (i) a shaded area, corresponding to a dome in the sphere, whereby any satellite has a distance from the source node that is smaller than $r_\mathrm{max}$ and can thus establish a connection; (ii) the remaining area, containing satellites that are too far from the source node to provide connectivity. 
As illustrated in Fig.~\ref{fig:sketch}, the distance $r_\mathrm{max}$ is related to the node-centered zenith angle $\varphi_{ \mathrm{s} }$ and the Earth-centered zenith angle $ \varphi_{ \mathrm{e} }$.
More precisely, $r_\mathrm{max}$ can be expressed as a function of $ \varphi_{ \mathrm{e} }$ and $\varphi_{ \mathrm{s} }$ as follows
\begin{equation}
    r_\mathrm{max} = \frac{(R_\oplus+h)\cos({ \varphi_{ \mathrm{e} })}-R_\oplus}{\cos({\varphi_{ \mathrm{s} }})},
\end{equation}
where $ \varphi_{ \mathrm{e} }$ is related to $\varphi_{ \mathrm{s} }$ as 
\begin{align} \
     \varphi_{ \mathrm{e} } = \cot^{-1}\left(\frac{\cot(\varphi_{ \mathrm{s}})+\rho\sqrt{\cot^2(\varphi_{ \mathrm{s}})+1-\rho^2}}{1-\rho^2}\right),
\end{align}
in which $\rho = R_\oplus/\left(R_\oplus+h\right)$.

As shown in Fig.~\ref{fig:sketch}, each satellite lying in the shaded dome region has a relative velocity with respect to the source node owed to its orbital movement and the Earth's rotation. Such relative speed causes the satellite to rapidly pass through the shaded region. 
In what follows, we focus on the satellites within the dome region and their motion mode. While there are multiple satellite orbits, the maximum arc length within the specific dome region remains fixed. We assume that the velocity difference between the satellites and the source node remains constant. 
By considering relative motion, we can treat all satellites within the dome region as stationary in space, similar to stars in the sky that appear static for a short period of time. Meanwhile, the source node undergoes circular movements at a velocity $v$. Furthermore, we assume that after completing a full cycle, the positions of the satellites are regenerated independently, following a homogeneous PPP with a density of $\lambda$. 

The duration for which a satellite remains within the dome region can be approximated as an on-service period, while the time between a satellite's departure from the dome region and the arrival of the subsequent satellite can be viewed as an off-service period. Due to the random positions of the satellites, the durations of both the on- and off-service periods vary. However, we can approximate these periods as independently and identically distributed. 
In the subsequent analysis, we leverage the techniques developed in \cite{madadi2017shared} to derive the distributions of the on- and off-service periods, taking into account the aforementioned approximations.

\begin{theorem}
\textit{The off-service periods follow an identical exponential distribution with density given by
\begin{align}
    \lambda_{ \mathrm{os}} = 2 \,\omega\,\lambda\,\sin (\varphi_{ \mathrm{e}}) \,\left(R_\oplus+h\right)^2
\end{align}
where $\omega = v/R_\oplus$ and the on-service periods have a common probability density function (PDF) given as
% On-Off process is the same as the serving process of an  $M/GI/\infty$ queue with arrival rate 
% $ \lambda_{ \mathrm{o} } = 2\left(R_\oplus+h\right)^2\omega\lambda\sin\varphi$ and i.i.d. service time with density
\begin{align}\label{equ:pdf}
f_W(s) =
\begin{cases}
    \frac{\omega\cos(\varphi_{ \mathrm{e}})\tan(\frac{\omega s}{2})}{2 \varphi_{ \mathrm{e} }\sqrt{\sin^2 (\varphi_{ \mathrm{e} })-\sin^2(\frac{\omega s}{2}})}, & \text{if }s \in \left[0,\frac{2 \varphi_{ \mathrm{e} }}{\omega}\right],  \\
    0, & \text{otherwise}.
\end{cases}
\end{align}
% where $\omega = v/R_\oplus$.
}
\end{theorem}
\begin{IEEEproof}
% Under our assumption, all satellites run similarly, keeping the same orbital inclination $0^\circ$. Thus, we can change the view to all satellites remain stationary in space and 
% the user keeps circular movements at velocity $v$. 
Note that the dome area changes as the source node moves. 
Since the satellites are scattered as a PPP with intensity $\lambda$, 
for a small value of time duration $\epsilon$, 
the number of satellites arriving for each period of duration $\epsilon$ is $\lambda\,A_\epsilon$, where 
\begin{equation} \label{equ:A_epsilon}
A_\epsilon = 2\left(R_\oplus+h\right)^2\,\omega\,\epsilon\,\sin (\varphi_{ \mathrm{e}})
\end{equation}
is the area of the dome moving in the interval $\epsilon$ as shown in Fig.~\ref{fig:user}. 
This implies that ($a$) the increments in the satellite arrival process have the same distribution (because \eqref{equ:A_epsilon} holds irrespective of the source node's position) and ($b$) the number of satellites in any two disjoint areas are independent.
% (For the characteristic of Poisson distribution, the number of nodes in two disjoint closed sets are independent, and for any $\epsilon>0$, the increments have the same distribution. Thus, the probability of having a single satellite arrive within the small interval $\epsilon$ is $\lambda \, A_\epsilon + o\left(\epsilon\right)$.)  
% \red{The probability of having a single satellite arrival within the interval $\epsilon$ is $\lambda \, A_\epsilon + o\left(\epsilon\right)$, while $\epsilon$ denotes a small value under the assumption of large satellite density.} 
As such, we can conclude that the off-service periods are exponentially distributed with the rate $2 \omega\lambda\sin (\varphi_{ \mathrm{e} }) \left(R_\oplus+h\right)^2$.

On the other hand, the trajectory of each satellite in the dome section is characterized by the latitude difference between the source node and the point of the satellite projected on the ground, where these together determine the total service time of each satellite when establishing a connection with the source node. 
Let $\Theta$ denote the angle difference that represents the possible entry location of satellites, following a uniform distribution on [$- \varphi_{ \mathrm{e} }$, $ \varphi_{ \mathrm{e} }$]. 
% ($\Theta$ represent the latitude difference, and is uniform on [$- \varphi_{ \mathrm{e} }$, $ \varphi_{ \mathrm{e} }$] , which bounds are decided by the dome region.)
% \red{Let $\Theta$ denote the angle difference and is uniform on [$- \varphi_{ \mathrm{e} }$, $ \varphi_{ \mathrm{e} }$] to describe the possible entry location of satellites.} 
The service time $W$ is thus a random variable given by
\begin{align} 
    W = \frac{2}{\omega}\arcsin\left( \frac{\sqrt{\sin^2 (\varphi_{\mathrm{e}})-\sin^2(\Theta)}}{\cos(\Theta)} \right).
\end{align}
The PDF of $W$ can then be derived using the distribution of $\Theta$.
% and whose density is provided in \eqref{equ:pdf}.
\end{IEEEproof}

%%%%%%%%%%%%%%%%%%%%%%%%%%%%%%%%%%%%%%%%%%%%%%

\subsection{Analytical Expression of the AoI in NTN}
% Without loss of generality, we focus on the $k$-th status update generation and reception time point. 
As illustrated in Fig.~\ref{fig:AoI}, we use $t_k$ and $t'_k$ to represent the timestamps corresponding to the transmit and receive instances of the $k$-th update.
We denote by $T_{o,i}$ and $T_{f,i}$ the duration of the $i$-th on- and off-service periods, respectively. 
If the $k$-th packet transmission is successful, $T_k = t'_k-t_k = D$ denotes the propagation delay and  
$Y_k = T_k+B_k$ indicates the period between the $k$-th and $k+1$-th received updates, where $B_k = t_{k+1}-t'_k$. 
% Using $T_k$ and $Y_k$, 
Following \cite{sinha2022age}, the time-average AoI can be expressed as
\begin{align}\label{equ:aoi_1}
    \bar{\Delta} &= \frac{\mathbb{E}\left[Y_{k}^{2}\right]+2\mathbb{E}\left[T_{k}Y_{k}\right]}{2\mathbb{E}\left[Y_{k}\right]}
    = \frac{\mathbb{E}\left[Y_{k}^{2}\right] }{2\mathbb{E}\left[Y_{k}\right]} + D.
\end{align}

Since the AoI evolves across a series of on- and off-service periods, calculating the expectations in \eqref{equ:aoi_1} entails determining two conditional probabilities : ($a$) the probability that the next update is generated during an on-service period, given that the latest update occurred in an on-service period, and ($b$) the probability that the next update will be generated in an off-service period if the latest update occurred during an off-service period. 
We denote these probabilities as $\mathrm{P}_{\mathrm{o}|\mathrm{o}}$ and $\mathrm{P}_{\mathrm{f}|\mathrm{f}}$, respectively, and we characterize them in the following lemma.
% Before deriving the moments of $Y_k$, we first need to calculate two conditional probabilities, one is the probability the next arrival is in $\mathrm{on}$ periods when the last arrival was in $\mathrm{on}$ periods, defined as 
% $\mathrm{P}_{(\mathrm{o}_1\mathrm{o}_2|\mathrm{o}_1)}$, and the other is the probability the next arrival is in $\mathrm{off}$ periods when the last arrival was in $\mathrm{off}$ periods, defined as $\mathrm{P}_{(\mathrm{f}_1\mathrm{f}_2|\mathrm{f}_1)}$. 

\begin{lemma}
\textit{
% The probability that the next upload is in the $\mathrm{On}$ state given that the last upload was in the $\mathrm{On}$ state and similarly, the probability that the next upload is in the $\mathrm{Off}$ state given that the last upload was in the $\mathrm{Off}$ state are
The conditional probability $\mathrm{P}_{\mathrm{f}|\mathrm{f}}$ is given by
\begin{align} 
    \mathrm{P}_{\mathrm{f}|\mathrm{f}} = \frac{1-a}{1-a\,b},
\end{align}
where $a = \lambda_{ \mathrm{os} }/\left(\mu+ \lambda_{ \mathrm{os} }\right)$ and $b =\int_{-\infty}^{\infty} e^{-\mu s}f_W(s)ds$.
The conditional probability $\mathrm{P}_{\mathrm{o}|\mathrm{o}}$ is
\begin{align} 
    \mathrm{P}_{\mathrm{o}|\mathrm{o}} = \frac{1-\left(2-\mathrm{P}_{\mathrm{f}|\mathrm{f}}\right)\mathrm{P}_\mathrm{off}}{1-\mathrm{P}_\mathrm{off}},
\end{align}
}
where $\mathrm{P}_\mathrm{off} = 1/\left(1+ \lambda_{ \mathrm{os} }\,\mathbb{E}\left[W\right]\right)$. 
\end{lemma}
\begin{IEEEproof}
% See Appendix~\ref{pro:Lem1} for a sketch of the proof.
% Based on the definition of $\mathrm{P}_{\mathrm{f}|\mathrm{f}}$, 
Without loss of generality, 
we assume an update is generated in the first off period at $t = 0$. 
We use $s'_i$ and $s_i$ to represent the start times of the $i$-th on- and off-service periods, respectively.
% To represent the start times of the $i$-th on- and off-service periods, we use $s'_i$ and $s_i$ respectively. 
As each on- and off-service period is distinct and non-overlapping, the probability of the next update remaining in an off-service period is obtained by adding up the probabilities of arrival within each individual off-service period:
\begin{equation}
\begin{aligned} 
& \mathrm{P_{\mathrm{f}|\mathrm{f}}^{\mathrm{Cond}}} = \textstyle\sum_{i=1}^{\infty} \left[ F\left(s'_i\right)-F\left(s_i\right) \right],\\
    &= \textstyle\sum_{i=1}^{\infty} \left[ \exp\left(-\mu s_i\right)-\exp\left(-\mu\left(s_i+T_{f,i}\right)\right) \right],\\
    &= \textstyle\sum_{i=1}^{\infty}\left[1-\exp\left(-\mu T_{f,i}\right)\right]\textstyle \prod_{k=1}^{i-1}\exp\left(-\mu\left(T_{f,k}+T_{o,k}\right)\right).
\end{aligned}
\end{equation}
By deconditioning random variables $\{ T_{o,i} \}_{i\geq1}$ and $\{ T_{f,i} \}_{ i \geq 1}$ in the above (and notice that they are i.i.d.), we have
\begin{equation}
\begin{aligned}
\mathrm{P_{\mathrm{f}|\mathrm{f}}} & =\textstyle\sum_{i=1}^{\infty}\mathbb{E}_{T_f}\left[1-e^{-\mu T_f}\right]\left(\mathbb{E}_{T_f}\left[e^{-\mu T_o}\right]\mathbb{E}_{T_o}\left[e^{-\mu T_f}\right]\right)^{i-1}\\
    & = \left(1-\frac{ \lambda_{ \mathrm{os} }}{\mu+ \lambda_{ \mathrm{os} }}\right)\sum_{l=0}^{\infty}\left(\frac{ \lambda_{ \mathrm{os} }}{\mu+ \lambda_{ \mathrm{os} }}\int e^{-\mu s}f_W(s)ds\right)^l\\
    & = \frac{1-\frac{ \lambda_{ \mathrm{os} }}{\mu+ \lambda_{ \mathrm{os} }}}{1-\frac{ \lambda_{ \mathrm{os} }}{\mu+ \lambda_{ \mathrm{os} }}\int e^{-\mu s}f_W(s)ds}.
\end{aligned}
\end{equation}
Let $\mathrm{P}_\mathrm{off}$ denote the probability that a random point is in an off-service period. We have that
\begin{align}
    \mathrm{P}_\mathrm{off} &= \frac{\sum_{l=0}^{\infty}\mathbb{E}[T_{f,l}]}{\sum_{l=0}^{\infty}(\mathbb{E}[T_{f,l}]+\mathbb{E}[T_{o,l}])} = \frac{1}{1+ \lambda_{ \mathrm{o} }\mathbb{E}[W]}
\end{align}
Let $\mathrm{P_{\mathrm{f}\mathrm{f}}}$ denote the probability that two adjacent updates are generated in one or different off-service periods and $\mathrm{P_{\mathrm{f}\mathrm{o}}}$ denote the probability that two adjacent updates that the previous is generated in an off-service period and the latter is generated in an on-service period. Similarly, we denote the probabilities of the other two cases as $\mathrm{P_{\mathrm{o}\mathrm{o}}}$ and $\mathrm{P_{\mathrm{o}\mathrm{f}}}$.
We obtain $\mathrm{P_{\mathrm{f}\mathrm{f}}} = \mathrm{P_{\mathrm{f}|\mathrm{f}}}\,\mathrm{P}_\mathrm{off}$ and $\mathrm{P_{\mathrm{f}\mathrm{o}}} = (1-\mathrm{P_{\mathrm{f}|\mathrm{f}}})\,\mathrm{P}_\mathrm{off}$. 
Since $\mathrm{P_{\mathrm{f}\mathrm{o}}} = \mathrm{P_{\mathrm{o}\mathrm{f}}}$, we have that
\begin{equation}
    \mathrm{P_{\mathrm{o}\mathrm{o}}} = 1 - \mathrm{P_{\mathrm{o}\mathrm{f}}} - \mathrm{P_{\mathrm{f}\mathrm{o}}} - \mathrm{P_{\mathrm{f}\mathrm{f}}} = 1-(2-\mathrm{P_{\mathrm{f}|\mathrm{f}}})\mathrm{P}_\mathrm{off}
\end{equation}
and obtain
\begin{align}
    \mathrm{P_{\mathrm{o}|\mathrm{o}}} = \frac{\mathrm{P_{\mathrm{o}\mathrm{o}}}}{1-\mathrm{P}_\mathrm{off}}.
\end{align}
\end{IEEEproof}

Using the results in Lemma~1, we can compute the first and second moments of $Y_k$ as follows.

\begin{lemma}\label{Lem2}
\textit{
The first and second moments of $Y_k$ are given by
\begin{align}\label{equ:yk}
    \mathbb{E}\left[Y_{k}\right] &= \frac{1}{\mu}+\frac{\gamma}{\mu}\left(1-\mathrm{P}_{\mathrm{o}|\mathrm{o}}\right), 
\end{align}\
and
\begin{align}\label{equ:yk_2} 
 \mathbb{E}\left[Y_{k}^{2}\right] &= \frac{2}{\mu^2}+\frac{2\left(\gamma+\gamma^2\right)}{\mu^2} 
    \left(1-\mathrm{P}_{\mathrm{o}|\mathrm{o}}\right),
\end{align}
where $\gamma= 1/\left(1-\mathrm{P}_{\mathrm{f}|\mathrm{f}}\right)$.
}
\end{lemma} 
\begin{IEEEproof}
% See Appendix~\ref{pro:Lem2} for a sketch of the proof.
Let $Z$ represent the interval time between two consecutive arrivals of status updates. Each update encounters one of the situations: ($a$) being accepted during the on-service period, or ($b$) being dropped during the off-service period. 
After a delivery failure, the AoI keeps increasing until a new arrival falls within the on-service period. 
By assuming that the subsequent arrival occurs during the on-service period when the previous one is accepted, we can easily derive
\begin{equation}
    \mathbb{E}\left[Y_{k}^{o}\right] = \mathbb{E}\left[Z\right] = \frac{1}{\mu}.
\end{equation}
Similarly, we obtain that
\begin{equation}
\begin{aligned}
    \mathbb{E}\left[Y_{k}^{f}\right] &= \sum_{n=2}^{\infty}\left[\sum_{l=1}^{n}\mathbb{E}\left[Z_l\right]\right]\mathrm{P_{\mathrm{f}|\mathrm{f}}}^{n-2}\left(1-\mathrm{P_{\mathrm{f}|\mathrm{f}}}\right),\\
    &= \frac{1}{\mu}\left(1+\frac{1}{1-\mathrm{P_{\mathrm{f}|\mathrm{f}}}}\right).
\end{aligned}
\end{equation}
The mean of $Y_k$ is thus given by
\begin{equation}
\mathbb{E}\left[Y_{k}\right]=\mathbb{E}\left[Y_{k}^{o}\right]\mathrm{P_{\mathrm{o}|\mathrm{o}}}+\mathbb{E}\left[Y_{k}^{f}\right](1-\mathrm{P_{\mathrm{o}|\mathrm{o}}}).
\end{equation}

Further, we can obtain the two terms forming the second moment of $Y_k$ respectively as
\begin{align}
    \mathbb{E}\left[Y_{k}^{2,o}\right] = \mathbb{E}\left[Z^2\right] = \frac{2}{\mu^2}
\end{align}
and
\begin{equation}
    \mathbb{E}\left[Y_{k}^{2,f}\right] = \sum_{n=2}^{\infty}\left[\left(\sum_{l=1}^{n}\mathbb{E}\left[Z_l\right]\right)^2\right]\mathrm{P_{\mathrm{f}|\mathrm{f}}}^{n-2}\left(1-\mathrm{P_{\mathrm{f}|\mathrm{f}}}\right).
\end{equation}
Since $Z_l$ follows exponential distribution with parameter $\mu$, we have $S_n = \textstyle\sum_{l=1}^{\infty}Z_l \thicksim \Gamma(n,\mu^{-1})$ and we obtain
\begin{equation}
\begin{aligned}
    \mathbb{E}\left[Y_{k}^{2,f}\right] &= \sum_{n=2}^{\infty}\left[\frac{n(n+1)}{\mu^2}\right]\mathrm{P_{\mathrm{f}|\mathrm{f}}}^{n-2}\left(1-\mathrm{P_{\mathrm{f}|\mathrm{f}}}\right),\\
    &= \frac{2}{\lambda^2}\frac{\mathrm{P_{\mathrm{f}|\mathrm{f}}^2}-3\mathrm{P_{\mathrm{f}|\mathrm{f}}}+3}{(1-\mathrm{P_{\mathrm{f}|\mathrm{f}}})^2}.
\end{aligned}
\end{equation}
The second moment of $Y_k$ thus yields
\begin{equation}
\mathbb{E}\left[Y_{k}^{2}\right]=\mathbb{E}\left[Y_{k}^{2,o}\right]\mathrm{P_{\mathrm{o}|\mathrm{o}}}+\mathbb{E}\left[Y_{k}^{2,f}\right](1-\mathrm{P_{\mathrm{o}|\mathrm{o}})}.
\end{equation}
\end{IEEEproof}

% By substituting the above into \eqref{equ:aoi_1}, we obtain the analytical expression of the time average AoI. 
Using the above lemmas, we can obtain the analytical expression of the time average AoI.
\begin{theorem}
\textit{The time-average AoI of the source-destination link in the considered system is
\begin{align} 
    % \bar{\Delta} = \frac{1+(\gamma+\gamma^2)(1-\mathrm{P}_{\mathrm{o}|\mathrm{o}})}{\mu+\mu\gamma(1-\mathrm{P}_{\mathrm{o}|\mathrm{o}})}+D.
    \bar{\Delta} = \frac{ \gamma^2 (1-\mathrm{P}_{\mathrm{o}|\mathrm{o}})}{\mu+\mu\gamma(1-\mathrm{P}_{\mathrm{o}|\mathrm{o}})} + \frac{1}{\mu} + D.
    \label{eq:AoI}
\end{align}
}
\end{theorem}
\begin{IEEEproof}
% From \eqref{equ:aoi_1}, the mean age can be written as 
% \begin{align} 
%     \bar{\Delta} = \frac{\mathbb{E}\left[Y_{k}^{2}\right]+2\mathbb{E}\left[T_{k}\right]\mathbb{E}\left[Y_{k}\right]}{2\mathbb{E}\left[Y_{k}\right]}.
% \end{align}
Equation (\ref{eq:AoI}) is obtained by substituting $\mathbb{E}[Y_{k}]$ and $\mathbb{E}[Y_{k}^2]$ from Lemma~\ref{Lem2} into \eqref{equ:aoi_1}.
% Substituting $\mathbb{E}\left[Y_{k}\right] = \bar{\mathrm{Y}}_{\mathrm{k}}^{1}$ , $\mathbb{E}\left[Y_{k}^2\right] = \bar{\mathrm{Y}}_{\mathrm{k}}^{2}$ from Lemma \ref{Lem2}\, and $\mathbb{E}\left[T_{k}\right] = D$.
\end{IEEEproof}

% ============================================ %
%         Section: Sim & Num Analysis          %
% ============================================ %
\section{Validation and Case Studies}

In this section, we validate the accuracy of our analysis and evaluate the time-average AoI under different NTN scenarios. 
For ease of notation, we employ the Earth-centered zenith angle $\varphi_{ \mathrm{s} }$, rather than $r_\mathrm{max}$, to represent the source node dome area. Unless otherwise specified, we use the following parameters: $h = 800\,\mathrm{km}$, $\omega = \pi/3600\,\mathrm{rad/s}$, $\varphi_{ \mathrm{s} } = 1^\circ$, and $D = 1\,\mathrm{s}$.

We evaluate the time-average AoI under a varying number of satellites in the constellation and for different status update arrival rates. 
Specifically, we deploy the satellite nodes on the sphere according to a homogeneous PPP. 
The source node cycles under velocity $v = \omega R_\oplus$.
Every time upon finishing a complete circle, we regenerate the positions of all satellites using the same distribution. 
We run each simulation with $10^{6}$ arrivals and collect the corresponding AoI statistics to calculate the average.

\begin{figure}[t!] 
  \centering{}
    {\includegraphics[width=\figwidth]{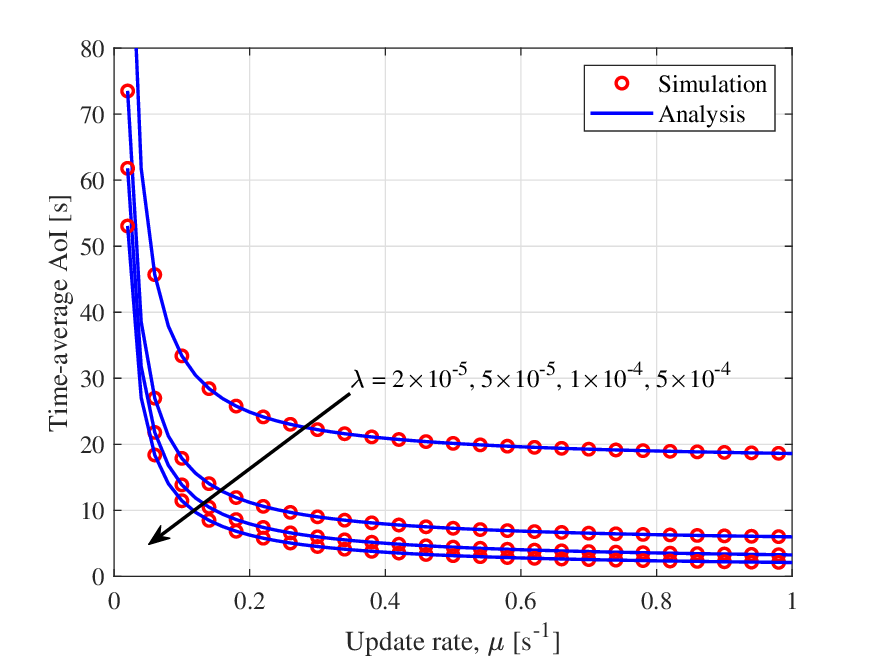}}
  \caption{Time-average AoI versus the status update rate under constellation densities $\lambda$ ranging from $2\times10^{-5}$ to $5\times10^{-4} \, \mathrm{km}^{-2}$, corresponding to $12924$ and $323100$ satellites deployed in the spherical surface, respectively.}
  \label{fig:sim_ana}
\end{figure}

% (Then we randomly distribute the points on the sphere to represent the satellites and set one position of the source node to simulate the NTN.
% For each time, the source node wraps around the sphere once, we regenerate the position of all random points on the sphere to ensure we have enough on- and off-service periods.
% For each simulation, we generate $10^{6}$ arrivals to obtain the time-average AoI compared with our analysis value.)
% \red{When the source node rotates one circle, we regenerate the position of all random points on the sphere. Thus we can simulate the on-off process by calculating the time point of satellites in and out of the dome area. For each arrival time, we check which on- or off-service periods it is in, and adopt different strategies to calculate AoI. We run each simulation with $10^{6}$ arrivals and collect the average AoI.}

Fig.~\ref{fig:sim_ana} shows the time-average AoI as a function of the source node status update rate, for different values of the total number of satellites passing the dome region in the constellation ranging from 225 to 5638.%
\footnote{As a reference, at the time of writing Starlink and OneWeb constellations respectively consist of about 4,000 and 500 LEO satellites \cite{SpaceX,SpaceX_Wiki,OneWeb_Wiki}.}
This figure shows a close match between the numerical results and our theoretical analysis, thus verifying the accuracy of Theorem~2. 
Additionally, Fig.~\ref{fig:sim_ana} shows that 
as deployment density increases, the rate of the time-average AoI descent gradually decreases. This is because the mean off-service period becomes smaller and the proportion of the on-service period becomes larger. It makes arrivals more probability get into the on-service period and be close to the value of $1/\mu$.

% \red{when the deployment density is low, there is an obvious contrast as it changes. As the deployment density increases to a certain extent, the mean AoI changes very little and is more likely to be decided by the arrival rate $\lambda$.}

\begin{figure}[t!] 
  \centering{}
    {\includegraphics[width=\figwidth]{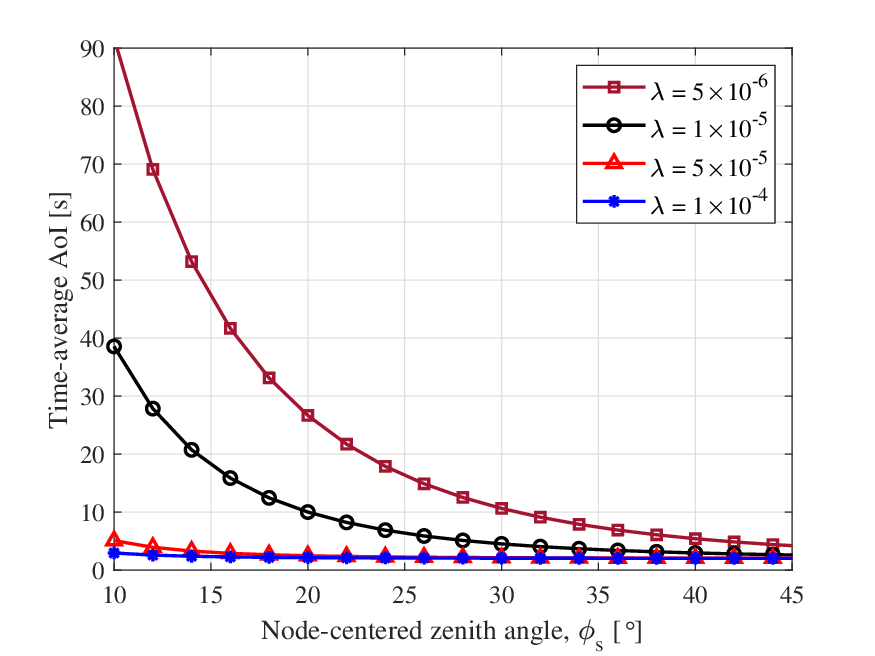}}
  \caption{Time-average AoI versus the node-centered zenith angle under constellation densities $\lambda$ ranging from $5\times10^{-6}$ to $1\times10^{-4} \, \mathrm{km}^{-2}$, corresponding to 3231 and 64620 satellites deployed in the spherical surface, respectively.}
  \vspace{-5mm}
  \label{fig:angle}
\end{figure}

Fig.~\ref{fig:angle} shows the time-average AoI as a function of the node-centered zenith angle. We note that the duration of the on-service periods increases as the node-centered zenith angle becomes larger. This causes the arrivals to have a higher probability of falling within on-service periods, and in turn decreases the resulting time-average AoI.

% ============================================ %
%         Section: Conclusion                  %
% ============================================ %
\section{Conclusion}

In this paper, we conducted a comprehensive analysis of the Age of Information in NTN, focusing on LEO satellite networks. By formulating an on-off process to approximate the service process between LEO satellites and a source node on Earth, we derived a closed-form expression for the time-average AoI using stochastic geometry tools. 
We also provided numerical results that validate the accuracy of our analysis and contribute to a deeper understanding of the factors influencing the AoI in NTN. Specifically, we quantified the impact of source status update rate and satellite constellation density on the time-average AoI. 
Overall, our study fills a gap in the literature by providing the first analysis of AoI in NTN and paves the way for future research in this timely field. 
\section*{Acknowledments}
The work of Y. Lu and H. H. Yang has been  supported in part by the Zhejiang Provincial Natural Science Foundation of China under Grant LGJ22F010001 and in part by the Zhejiang Lab Open Research Project (No. K2022PD0AB05).
The work of N. Pappas has been supported in part by the Swedish Research Council (VR), ELLIIT, Zenith, and the European Union (ETHER, 101096526).
The work of G. Geraci was supported in part by the Spanish Research Agency through grants PID2021-123999OB-I00 and CEX2021-001195-M, by the UPF-Fractus Chair, and by the Spanish Ministry of Economic Affairs and Digital Transformation and the European Union through grants TSI-063000-2021-59 (RISC-6G), TSI-0630002021-138 (6G-SORUS), and TSI-063000-2021-52 (AEON-ZERO).

%%%%%%%%%%%%%%%%%%%%%%%%%%%%%%%%%%%%%%%%%%%%%%%%%%%%

\bibliographystyle{IEEEtran}
\bibliography{SupportDocuments/bib/IEEEabrv,SupportDocuments/bib/StringDefinitions,SupportDocuments/bib/howard_AoI_Ctrl}

\end{document}